# Design of a semiconductor ferromagnet in a quantum dot artificial crystal

Kenji Shiraishi[a)], Hiroyuki Tamura[b)], Hideaki Takayanagi

*NTT Basic Research Laboratories, Atsugi, Kanagawa 243-0198, Japan*

We present the theoretical design of quantum dot (QD) artificial ferromagnetic crystals. The electronic structure calculations based on local spin density approximation (LSDA) show that our designed QD artificial crystal from a structure comprising the crossing 0.104-µm-wide InAs quantum wires (an effective Kagome lattice) has flat band characteristics. Our examined QD artificial crystal has the ferromagnetic ground state when the flat band is half-filled, even though it contains no magnetic elements. The ferromagnetic and the paramagnetic state can be freely switched by changing the electron filling via a gate voltage.

---

[a)] Present address: Institute of Physics, University of Tsukuba, Tsukuba, Ibaraki 305-2635, Japan
[b)] Author to whom all correspondence should be addressed; electronic mail: tamura@will.brl.ntt.co.jp

A quantum dot (QD) artificial atom is a tiny fabricated region in semiconductors, where electrons are localized by a confinement potential.[1] It has two crucial characteristics that are very different from those of a real atom; the spatial position is fixed because a QD artificial atom is a rigid buried region in semiconductors, and the number of electrons in it can be changed in a controllable manner. These characteristics provide us a novel concept of a QD artificial crystal in material science: By treating a QD artificial atom as a building block, we can design a QD artificial crystal with any structure as we like regardless of the number of electrons in it. In a real atom, however, such unrestricted material design is impossible, since valence electrons also govern the crystal structure. We demonstrate the theoretical design of QD artificial ferromagnetic crystals made up of nonmagnetic QD artificial atoms.

QD artificial atoms have well defined quantum states and a variable number of electrons (from one to a thousand). Thus, they provide a special stage for the examination of the basic principles of quantum mechanics. By investigating the energy-level splitting under an applied magnetic field, characteristic shell structures corresponding to circular QD artificial atoms have been observed[2] where the shell filling obeys Hund's rule.

By treating a QD artificial atom as a building block, new types of molecules (QD artificial molecules) can be created. Recently, two connected QD artificial atoms (QD artificial molecules) were found to have "ionic" or "covalent" bonds, depending on the strength of inter-dot coupling.[3] QD artificial molecules have also been suggested as promising candidates for quantum computing devices.[4]

QD artificial crystals may even be possible if a large number of QD artificial atoms can be connected to each other.[5-7] We will be able to create QD artificial crystals whose structures are impossible in real crystals: In real crystals, the number of valence electrons governs the structure. For example, in a diamond structure, each chemical bond contains two electrons. Only atoms of group-IV elements such as Si can form stable diamond structures, whereas atoms of group-III elements cannot; Al only forms a face-centered-cubic (f.c.c.) structure. However, QD artificial atoms enable us to fabricate any type of crystal structure regardless of the number of valence electrons in them. This is because the QD artificial crystals, whose building blocks (QD artificial atoms) are buried regions rigidly fabricated in the host semiconductor, do not undergo structural deformations by electronic effects such as Jahn-Teller distortion.

Naturally, therefore, the QD artificial crystal concept would allow us to design any type of physical property originating from the characteristic crystal structure without having to consider the structural stability, which is crucial in real crystals. In other words, mathematical lattice models that have so far been discussed only in theoretical physics can be realized. Based on the mathematical modeling of the Hubbard Hamiltonian, some works have stressed the importance of crystal structures in the appearance of certain important physical properties.[8-13] For example, Lieb[8], Mielke[9], and Tasaki[10] have indicated that ferromagnetism appears in certain lattice models when the flat band is half-filled. However, it is difficult to form structures with properties of these lattices using conventional materials because Jahn-Teller distortion would destabilize the structures when the flat-band is half-filled, although some proposals have been made.[14-16] The eventual fabrication of QD artificial crystal materials with such striking properties is the goal of our research. In this study, we present a first-principles design of ferromagnetic materials made up of nonmagnetic QD artificial atoms. The structures we designed are realistic enough to be fabricated by existing nanotechnologies.

To design the electronic structures, we performed first-principles electronic structure calculations based on the local spin density functional approximation (LSDA) with the Perdew-Zunger exchange-correlation potential.[17] In the calculations, wave functions are expanded by a plane wave basis set.[18] The structure we examined contains crossing quantum wires with square cross sections as shown in Fig. 1. The width of each quantum wire is 0.104 μm and the lateral size of each two-dimensional unit cell is 0.72 μm. We assume that the quantum wire is InAs surrounded by $In_{0.72}Ga_{0.28}As$ barrier regions with barrier height of 0.21 eV. Since the effective width of quantum wires at the cross points is larger than the normal width of the wire, the cross points can act as QD artificial atoms. Accordingly, we expected that this artificial structure would prove to be effectively identical to a Kagome crystal, for which flat-band ferromagnetism has been predicted.[9]

First, we investigated the electronic structures of paramagnetic states, and examined whether a Kagome crystal is truly formed. The band structure of the paramagnetic configuration when five electrons are contained in each unit cell is given in Fig. 2(a). One can see that the third lowest band is dispersionless and half-filled. This flat-band characteristic is well reproduced by the tight-binding Hamiltonian of the Kagome lattice. This indicates that, as expected, the quantum-wire based structure acts as a Kagome crystal. This is also confirmed by the calculated total charge density. The calculated total charge density has maxima at the rectangularly shaped cross points, thus forming a Kagome crystal as shown in Fig. 2(b).

Next, we examined the ferromagnetic state of this Kagome crystal. The calculated spin density has maximums at the cross points, forming a Kagome crystal as shown in Fig. 3(a). This figure clearly indicates the appearance of ferromagnetism. The calculated band structures show that the Fermi level is located between up- and down-spin flat-bands, reflecting the ferromagnetism as shown in Fig. 3(b). The energy separation between up- and down-spin bands is 0.05 meV. The total energy of the ferromagnetic state is lower than that of the paramagnetic state by about 130 mK. These results indicate that the ground state of this Kagome crystal is ferromagnetic.[19] It is surprising that whether a non-magnetic semiconductor changes into a ferromagnetic material depends solely on the shape of the array of artificial atoms. We also found that the ground state is still ferromagnetic when the flat band is partly filled other than half-filled, and becomes paramagnetic when the flat band is empty or fully filled. Therefore, the ferromagnetic and the paramagnetic state can be freely switched when the electron filling in the flat band is modulated by applying a gate voltage. We also found that the structures with smaller wire width enhance the stability of ferromagnetic states; the energy difference between para- and ferromagnetic states in the effective Kagome crystal with 10 nm-wide InAs wires becomes 2.3 K.

Now, we comment on the feasibility of actual fabrications and measurements of our designed QD artificial crystals. An advantage of this quantum-wire-based structure is that it would be rather easy to realize by using existing nanofabrication techniques; Considering that even the formation of 8-nm-wide semiconductor quantum wires has been reported,[20] we can say that our designed structures with 0.104 μm-wide semiconductor wires could be fabricated by using existing electron beam (EB) lithography and etching techniques. It is also promising that potential control by the split-gates also enables the formation of QD artificial ferromagnetic crystals. In addition, it has been reported that small spins in mesoscopic systems can be detected, for instance, by conductance measurements.[21] For the above reasons, we expect to be able to fabricate a QD artificial ferromagnetic crystal and detect ferromagnetic spins in it.

Our finding can easily be expanded to include the design of artificial materials with other physical properties such as superconductivity. Superconductive properties have been predicted by mathematical models of many–body theory.[12,13] If we could design QD artificial crystals that reflect these mathematical models, they should possess superconductive properties. In addition, similar material design based on Si and GaAs nanostructures should be also possible. If Si-based QD artificial crystal revealed ferromagnetism, it would lead to two significant advantages. One is that magnetic devices could be fabricated on the same Si-based LSI chips. The other is that Si-based magnetic devices would not contain any toxic elements such as Cr. These advantages will lead to significant advances in information technologies and environmental cautious technologies, as well as in nanotechnologies.

We are grateful to Prof. H. Aoki, Prof. K. Kusakabe and Dr. R. Arita for their stimulating discussions. We also thank Dr. S. Ishihara for his continuous encouragement and helpful advice. This work was partly supported by JSPS under contract No. RFT96P00203 and the NEDO International Joint Research Grant.

**Figure Captions**

**Figure 1** The effective Kagome crystal we examined. The crossing quantum wire is InAs. The cross section of each wire is square shaped. The cross point array is identical to a Kagome crystal.

**Figure 2** Electronic structures of paramagnetic states calculated by using the local spin density approximation (LSDA). **(a)** LSDA-calculated paramagnetic band structures. The corresponding Brillouin zone is shown in the inset. **(b)** A cross-sectional view of LSDA-calculated total charge density of paramagnetic states.

**Figure 3** Electronic structures of ferromagnetic states calculated by using the local spin density approximation (LSDA). **(a)** A cross-sectional view of LSDA-calculated spin density of ferromagnetic states. The corresponding Brillouin zone is shown in the inset. **(b)** LSDA-calculated ferromagnetic band structures. Solid and dotted bands represent the up- and down-spin bands, respectively.

**Figures**

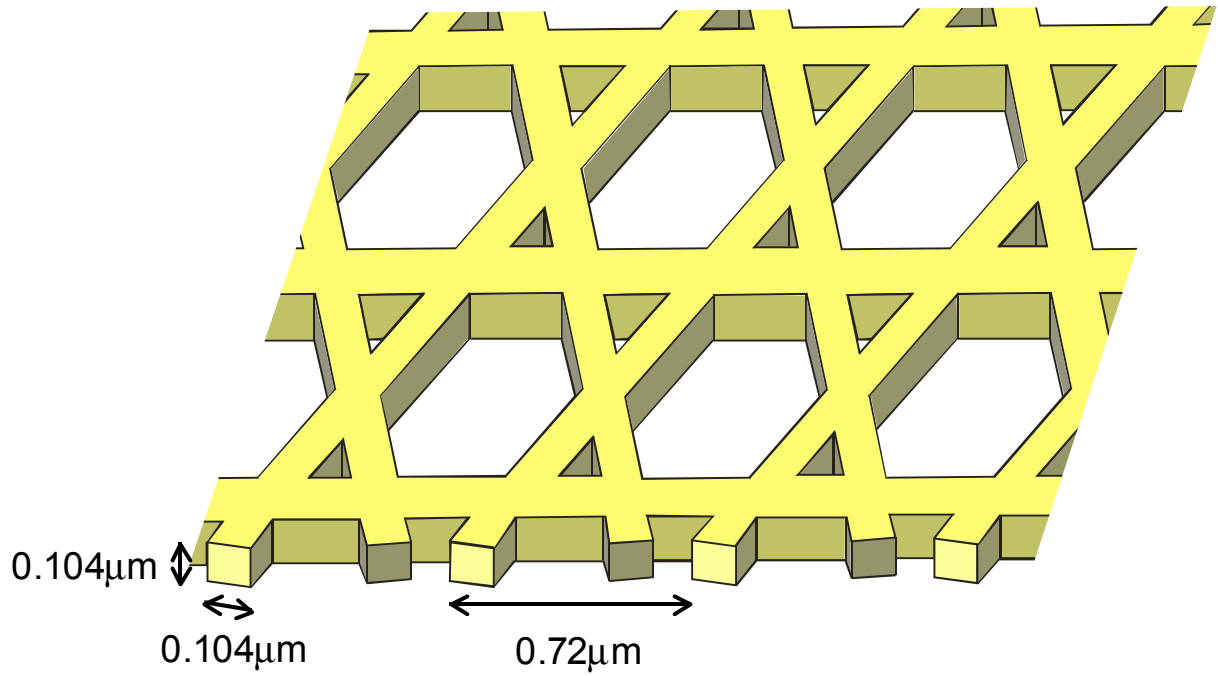

Fig. 1 K. Shiraishi *et al.*

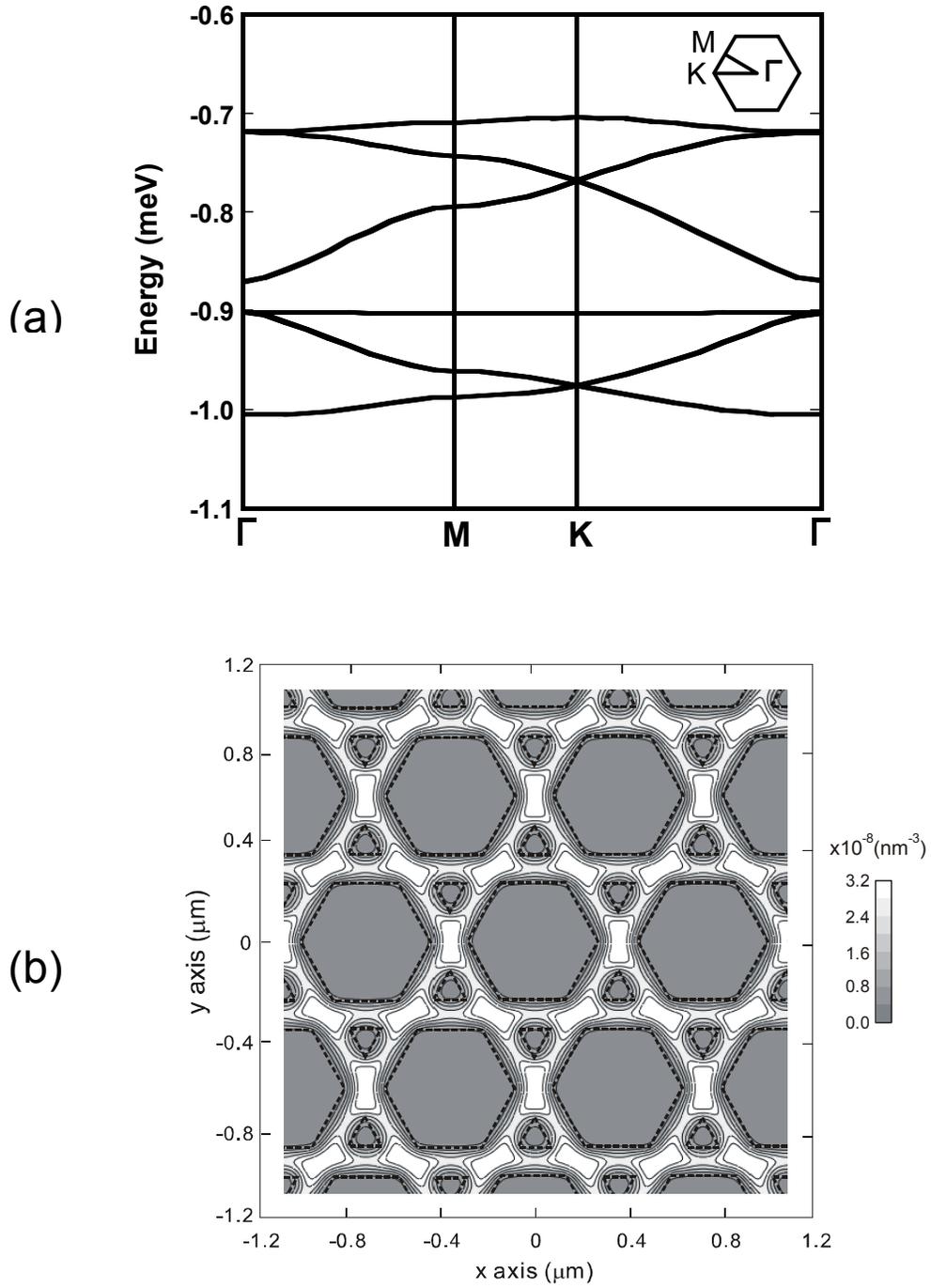

Fig. 2    K. Shiraishi *et al.*

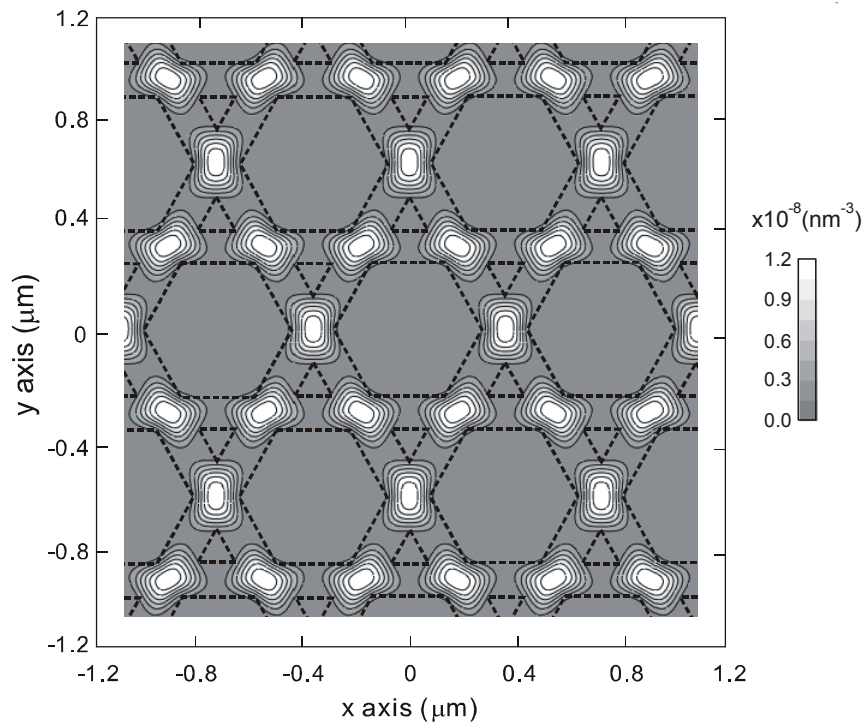

(b)

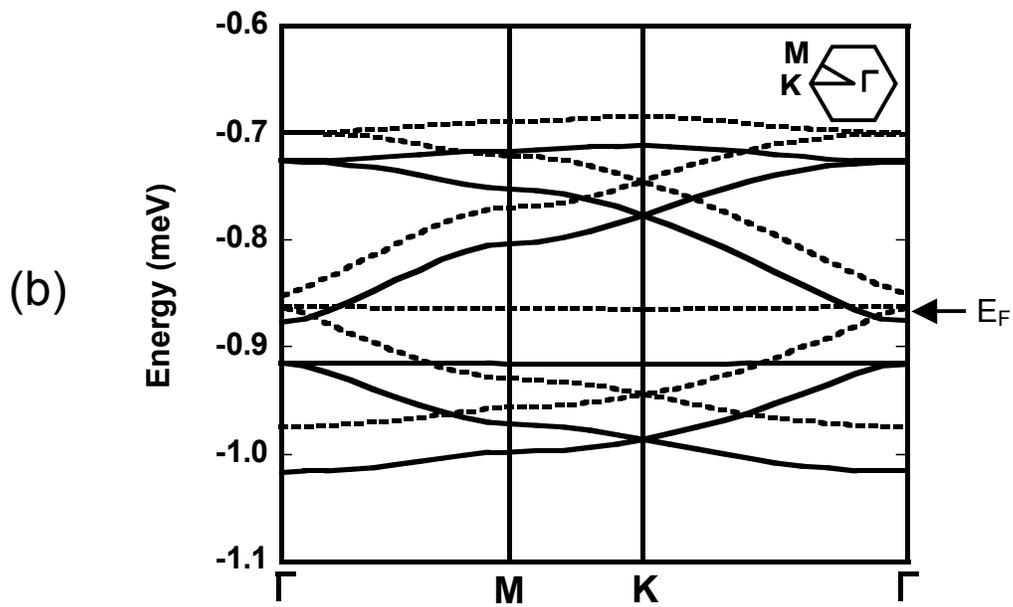

Fig. 3  K. Shiraishi *et al.*